\documentstyle[prb,aps,psfig,epsf]{revtex}

\begin{document}
\title{Josephson current in superconductor-ferromagnet structures with a
nonhomogeneous magnetization}
\author{F.S.Bergeret $^{1 }$, A.F. Volkov$^{1,2}$ and K.B.Efetov$^{1,3}$}
\address{$^{(1)}$Theoretische Physik III,\\
Ruhr-Universit\"{a}t Bochum, D-44780 Bochum, Germany\\
$^{(2)}$Institute of Radioengineering and Electronics of the Russian Academy%
\\
of Sciences, 103907 Moscow, Russia \\
$^{(3)}$L.D. Landau Institute for Theoretical Physics, 117940 Moscow, Russia }
\maketitle

\begin{abstract}
We calculate the dc Josephson current $I_J$ for two types of
superconductor-ferromagnet (S/F) Josephson junctions. The junction of the
first type is a S/F/S junction. On the basis of the Eilenberger equation, the
Josephson current is calculated for an arbitrary impurity concentration. If $%
h\tau\ll1$ the expression for the Josephson critical current $I_c$ is
reduced to that which can be obtained from the Usadel equation ($h$ is the
exchange energy, $\tau$ is the momentum relaxation time). In the opposite
limit $h\tau\gg1$ the superconducting condensate oscillates with period $%
v_F/h$ and penetrates into the F region over distances of the order of the
mean free path $l$. For this kind of junctions we also calculate $I_J$ in
the case when the F layer presents a nonhomogeneous (spiral) magnetic
structure with the period $2\pi /Q$. It is shown that for not too low
temperatures, the $\pi$-state which occurs in the case of a homogeneous
magnetization ($Q=0$) may disappear even at small values of $Q$. In this
nonhomogeneous case, the superconducting condensate has a nonzero triplet component and can
penetrate into the F layer over a long distance of the order of $\xi_{T}=%
\sqrt{D/2\pi T}$. The junction of the second type consists of two S/F
bilayers separated by a thin insulating film. It is shown that the critical
Josephson current $I_{c}$ depends on the relative orientation of the
effective exchange field $h$ of the bilayers. In the case of an antiparallel
orientation, $I_{c}$ increases with increasing $h$. We establish also that
in the F film deposited on a superconductor, the Meissner current created by
the internal magnetic field may be both diamagnetic or paramagnetic.
\end{abstract}

\section{Introduction}

Interplay between ferromagnetism and superconductivity in layered structures
has attracted a great interest in the last years. In a rough
approximation, these states are antagonistic to each other and the
ferromagnetism, being usually much stronger than superconductivity, is
supposed to destroy the latter.   However, in many cases the coexistence of these
two phenomena is possible, even if the superconducting critical temperature
$T_c$ is by order of magnitude lower than the Curie temperature of the
ferromagnet. Such is the case when dealing with superconductor-ferromagnet (S/F)
hybrid structures. In these systems the mutual interaction of these two states
may lead to significant changes of the thermodynamic and transport properties. 

In particular for  S/F/S systems in equilibrium, one of the
most interesting effects is a phase shift by $\pi $ between weakly coupled
superconductors, the so called $\pi $-state. The possibility of the $\pi $
-state in S/F/S structures was first predicted by Bulaevskii et al. \cite
{Bul,Buzdin} and studied in later works \cite{Clem,Buzdin1}.The
transition to the $\pi $-state manifests itself in a nonmonotonic (and even
oscillatory) thickness dependence either of the superconducting critical
temperature $T_{c}$ or of the critical current $I_{c}$, and in the change of
sign of $I_{c}$ if the exchange field $h$ exceeds a certain value in the
 S/F/S junction (Refs.\cite{Bul,Buzdin,Clem,Buzdin1,Jiang,Koor,Ryaz}). Although some
experiments on the thickness dependence of $T_{c}$ in S/F structures show
that for a certain thickness of the F layer, the ground state of the system
may correspond to the $\pi $-phase shift between the adjacent
superconductors (see e.g. Ref.\cite{Jiang}), this kind of coupling was not
observed in other experiments(see, for example, Ref. \cite{Koor}). Only
recently, the experiment of Ref. \cite{Ryaz} on the measurement of the
current $I_{c}(T)$ in the {\em ${\rm Nb}/{\rm Cu}_{x}{\rm Ni}_{1-x}/{\rm Nb}$%
} Josephson junction  demonstrated unambiguously the transition from the $0$
to the $\pi $-phase difference between the superconductors.

In all theoretical works \cite{Bul,Buzdin,Clem,Buzdin1},
calculations were performed either in the diffusive limit, in which the
Usadel equation was applicable, or in the pure ballistic limit, where the
elastic scattering by impurities was completely neglected. At the same time,
very often the parameters characterizing the samples in experiments, such as
the sample size, the mean free path or the strength of the exchange field,
do not correspond to these limits. Therefore, there is a certain need to
study the Josephson current in the S/F/S structures not only in
extreme limits but also in the intermediate region of the parameters.

In this work, we calculate the critical Josephson currents $I_{c}$ in a
S/F/S junction for arbitrary impurity concentrations. Since the approach
based on the Usadel equation \cite{Usadel} (dirty limit) is valid only if
the parameter $h\tau $ is small ($h$ is the exchange field of the
ferromagnet and $\tau $ is the momentum relaxation time), we use in an
arbitrary case  the more general Eilenberger equation \cite{Eilen,LOvchQuas}
in which, generally speaking, the elastic collision integral is not
neglected. As mentioned above, in real experiments the parameter $h\tau $
may take different values depending on the sample and therefore our theory can serve as a good description of the experiments.

Moreover, in all theoretical works mentioned previously, it was assumed that
the magnetic ordering in the ferromagnetic layers was homogeneous. However,
ferromagnetic materials  exhibit generally more complex magnetic structures.
In strong ferromagnets, like Fe or Ni, the magnetic ground state consists of
homogeneously magnetized domains with different relative orientations.
Also ``weak'' ferromagnets, like some ternary compounds with a regular
lattice of rare earths elements, turn out to be superconducting as the
crystals undergo a transition into a state with a nonhomogeneous
(helicoidal) magnetic order (see Ref. \cite{Advanc} and references
therein). A similar nonhomogeneous structure may arise in bilayered S/F
structures. For example, an experiment \cite{Muhge} and two theoretical
works \cite{Berger,Bul1} suggested a possible  existence of a
nonhomogeneous magnetic ordering in the ferromagnetic layer in a S/F system.
Also in experiments on giant magneto-resistance (GMR) in magnetic
multilayers employing superconducting contacts, nonhomogeneous magnetic
structures can be created artificially (see e.g. the review Ref.\cite{Bauer}
and references therein).

In spite of the importance for the experiments, a theoretical analysis of
the influence of a nonhomogeneous magnetization on the properties of S/F
junctions is still lacking. Therefore, the second goal of this paper is to
investigate the influence of  nonhomogeneous magnetic configurations on
the supercurrent through different kinds of superconductor-ferromagnet
Josephson junctions.

In Sec. \ref{section_sfs} we consider a S/F/S system, with an nonuniform
(spiral) magnetic ordering. We derive an expression for the critical current 
$I_{c}(Q)$ , where $Q$ is the wave vector of the spiral magnetic order. We
show that, whereas for $Q=0$ the transition from the $0$-phase state to the $%
\pi $-phase state is possible, even small nonzero $Q$ values may restore the 
$0$-phase state provided that the temperature is not too low. The reason for this
 is the existence of a triplet component of the superconducting condensate in
the ferromagnet due to the proximity effect and the nonhomogeneous magnetic
structure. In the limit $h\tau <1$ this component does not decay over the
short distance $\sqrt{D/h}$, which corresponds to the  length of decay of  the
usual singlet component, surviving up to a much longer distance $\sim \sqrt{%
D/2\pi T}$ ($D$ is the diffusion coefficient). The influence of this triplet
component on the transport properties of the S/F mesoscopic structures was
studied in Ref. \cite{BVE2}.

In Sec. \ref{section_sfisf}, we analyze the dc Josephson current in a tunnel
junction composed either of two S/F bilayers or of two magnetic
superconductors. We derive an expression for the critical current $I_{c}$
as a function of the relative angle $\alpha $ between the magnetization of
both F layers. The most important and surprising result is that for an
antiferromagnetic configuration, $\alpha =\pi $, the current $I_{c}$
increases with increasing  exchange field $h$. The calculated dependence
of $I_{c}$ on various parameters allows us to make some conclusions not only
on the magnetic order of the ferromagnetic materials used in S/F structures
but also on nonhomogeneous superconducting states predicted by Fulde,
Ferrel, Larkin and Ovchinnikov \cite{Fulde,LOv}.

In Sec.V we show that a Meissner current is induced in the F-region due to
the internal magnetic field of the ferromagnet. The Meissner current density
has a different sign at different points, and the total current in the
ferromagnet is either diamagnetic or paramagnetic depending on the thickness 
$d$ of the F film.

In the Appendix we present the  derivation of the main equations used in this
article. All our calculations are based on the
Eilenberger\cite{Eilen,LOvchQuas} or on the Usadel\cite{Usadel} equations, generalized to
the case of a spin-dependent interaction varying in space. Another approach
based mainly on the Bogolyubov-De Gennes equations was widely used for the study
of the spin-injection from a ferromagnet into unconventional superconductors
(see, e.g. Ref.\cite{Tanaka} and references therein). In the present work, we
restrict ourselves to the case of conventional superconductors with s-wave pairing.

\section{The Josephson current in a S/F/S structure}

In this section we calculate the dc Josephson current $I_{J}$ in a S/F/S
structure. In order to make the consideration as general as possible we use
the Eilenberger equation \cite{Eilen} including the elastic collision term.
This allows us to calculate $I_{J}$ for an arbitrary impurity concentration
and to formulate conditions under which the ballistic or diffusive limits
can be obtained. In order to find the condensate Green's function $\hat{f}%
_{\omega }$ in the F region in an analytical form, we assume that the
proximity effect is weak, {\it i.e.} $|\hat{f}_{\omega }|\ll 1$, and
linearize the collision term in the Eilenberger equation. This assumption
can be reasonable for structures with a big mismatch between the Fermi
surfaces in F and S, which leads to a small transmission coefficient $T$
through the S/F interface. If the coefficient $T$ is of the order of unity,
we hope that our results are valid at least qualitatively.



We consider the S/F/S structure shown in Fig. 1 and assume
that the exchange energy $h$ is homogeneous in the F region (the case of a
non-homogeneous $h$ will be analyzed in the next section). Because of the
small interface transparency one can neglect the suppression of the order
parameter $\Delta$ in the superconductor due to the proximity of the
ferromagnet. We assume also that there are no spin-flip processes in the
ferromagnetic region, {\it i.e.} the spin-relaxation length is larger than
the thickness $d$ of the ferromagnet and there are no spin-processes at the
S/F interface. The linearized Eilenberger equation in the Matsubara
representation has the following form 
\begin{equation}
\mu l\hat{\tau}_3\partial_x\hat{f}+2\left(\omega_m -ih\right)\hat{f}={\rm sgn}\omega\left(\langle\hat{f}\rangle-\hat{f}\right)\; .
\label{eillin}
\end{equation}

Here $\hat{\tau}_3$ is the Pauli matrix, $\mu =\cos\theta$, $\theta$ is the
angle between the momentum and the $x$-axis, $l=v_F\tau$ is the mean free
path, and $\omega_m=\pi T(2m+1)$ is the Matsubara frequency. The angle
brackets denote the average over angles: $<...>=(1/2)\int_{-1}^1{\rm \ d}%
\mu(...)$. Eq. (\ref{eillin}) is complemented by the boundary conditions at $%
x=\pm d/2$, which in the case of low transparency take the form \cite
{Zaitsev}

\begin{equation}
\hat{a}=(\gamma /2)\left[ {\rm sgn}\omega \hat{\tau}_{3}+\hat{s},\hat{g}_{s}+%
\hat{f}_{s}\right] \cong \gamma {\rm sgn}\omega \left( \hat{\tau}_{3}\hat{f}%
_{s}\right) _{x=\pm d/2}\;,  \label{bound_cond}
\end{equation}
where $\hat{a}$ and $\hat{s}$ are the antisymmetric and symmetric (with
respect to $\mu $) parts of $\hat{f}$, $\gamma =T(\mu )/4$ is a parameter
describing the transmittance of the interface, $T(\mu )$ is the transmission
coefficient, $\hat{g}_{s}$ and $\hat{f}_{s}$ are the quasiclassical normal
and anomalous Green's functions of the superconductors. The square brackets
denote the commutator. When writing the last equality, we neglect the term
proportional to $\hat{f}$, since $|\hat{f}|\sim \gamma $. The condensate function $\hat{%
f}_{s}$ in the superconductors can be written as  
\begin{equation}
\hat{f}_{s}(\pm d/2)=\left( i\hat{\tau}_{2}\cos (\varphi /2)\pm i\hat{\tau}%
_{1}\sin (\varphi /2)\right) f_{s}\;,  \label{eq:f_in_SC}
\end{equation}
where $f_{s}=\Delta /\sqrt{\Delta ^{2}+\omega _{m}^{2}}$ and $\varphi $ is
the  phase difference between the superconductors. 

It is convenient to represent the solution of Eq. (\ref{eillin}) as a sum of
the symmetric ($\hat{s}$) and the antisymmetric ($\hat{a}$) parts, 
\begin{equation}
\hat{f}=\hat{s}+\hat{a}\;  \label{separation}
\end{equation}
Substituting this expression into Eq. (\ref{eillin}) and separating the
symmetric and antisymmetric terms, one obtains two equations which determine
the functions $\hat{a}$ and $\hat{s}$: 
\begin{eqnarray}
\hat{a} &=&-{\rm sgn}\omega \left( \mu l/\kappa _{\omega }\right) \hat{\tau}%
_{3}\partial _{x}\hat{s}  \label{anti} \\
\mu ^{2}l^{2}\partial _{xx}^{2}\hat{s}-\kappa _{\omega }^{2}\hat{s}
&=&-\kappa _{\omega }\left\langle \hat{s}\right\rangle \;,  \label{symet}
\end{eqnarray}
where $\kappa _{\omega }=(1+2|\omega _{m}|\tau )-{\rm sgn}\omega 2ih\tau $.
Thus, the problem is reduced to finding the solution for Eq. (\ref{symet})
in the interval $|x|<d/2$ with the boundary conditions given by Eqs. (\ref
{bound_cond}) and (\ref{anti}). To this end it is convenient to extend
formally the function $\hat{s}$ over the whole $x-$axis and to write Eq. (%
\ref{symet}) in the following form 
\begin{equation}
\mu ^{2}l^{2}\partial _{xx}^{2}\hat{s}-\kappa _{\omega }^{2}\hat{s}=-\kappa
_{\omega }\left[ \langle \hat{s}\rangle +2\mu l\gamma f_{s}\sum_{n=-\infty
}^{\infty }\left( i\hat{\tau}_{2}\cos (\varphi /2)+(-1)^{n}i\hat{\tau}%
_{1}\sin (\varphi /2)\right) \delta \left( x-(d/2)(2n+1)\right) \right] \;.
\label{extent}
\end{equation}
One can prove that the solution of Eq. (\ref{extent}) obeys the boundary
conditions. 

Performing the Fourier transformation, we find the solution for the Fourier
transform of $\hat{s}$ 
\begin{equation}
\hat{s}_{k}=2f_{s}.B.\hat{F}\;,  \label{fourier_s}
\end{equation}
where 
\begin{eqnarray}
B &=&\frac{\kappa _{\omega }}{\left( 1-\kappa _{\omega }\langle
M^{-1}\rangle \right) M}\left[ \gamma \mu l-\kappa _{\omega }l\left( \mu
\gamma \langle M^{-1}\rangle -\langle \mu \gamma /M\rangle \right) \right]
\;,  \label{B} \\
\hat{F} &=&\sum_{-\infty }^{\infty }\exp (ikd(2n+1)/2)\left( i\hat{\tau}%
_{2}\cos (\varphi /2)+(-1)^{n}i\hat{\tau}_{1}\sin (\varphi /2)\right) \;
\label{F}
\end{eqnarray}
and 
\begin{equation}
M=(\mu lk)^{2}+\kappa _{\omega }^{2}\;.  \label{M}
\end{equation}
The function $\hat{s}$ determines the dc Josephson current, as well as the
variation of the density of states (DOS) due to the proximity effect. It is
given by the inverse Fourier transformation 
\begin{equation}
\hat{s}(x)=\int {\frac{{\rm d}k}{2\pi }\exp (-ikx)\hat{s}_{k}}\;.
\label{inv_fourier}
\end{equation}
The current is determined by the expression 
\begin{eqnarray}
I_{J} &=&\frac{1}{8}G_{Q}N(2\pi i)(2T){\rm Tr}\hat{\tau}_{3}\sum_{\omega
_{m}}{\langle a_{\omega }\mu \rangle }=  \label{current_sfs} \\
&&\frac{1}{8}G_{Q}N(2\pi i)(2T){\rm Tr}\hat{\tau}_{3}\sum_{\omega _{m}}{%
\left\langle (\mu \gamma /2)\left[ \hat{s}(d/2),f_{s}\right] \right\rangle }%
\;,
\end{eqnarray}
where $G_{Q}=e^{2}/\hbar $, $N=k_{F}^2{\cal S}/\pi ^{2}$ and ${\cal S}$ is the
cross-section area of the junction. In writing Eq. (\ref{current_sfs}) we
have used the boundary condition Eq. (\ref{bound_cond}). The summation over
Matsubara frequencies is carried out from $m=-\infty $ to $m=+\infty $.
Substituting Eqs. (\ref{symet}-\ref{inv_fourier}) into Eq. (\ref{current_sfs}%
), we obtain finally the dc Josephson current 
\begin{equation}
I_{J}=I_{c}\sin \varphi   \label{jos_current}
\end{equation}
where 
\begin{equation}
I_{c}=G_{Q}NJ_{c}\;,  \label{crit_current}
\end{equation}
and 
\begin{equation}
J_{c}=2\pi T{\rm Re}\sum_{\omega >0}^{\infty }{\sum_{n=0}^{\infty }{%
f_{s}^{2}\int {\frac{{\rm d}k}{2\pi }\langle \gamma \mu B\rangle \exp {%
\left( ikd(2n+1)\right) }}}}\;.  \label{crit_general}
\end{equation}
Eq. (\ref{crit_general}) determines the critical current and is valid for
any impurity concentration. Its analytical evaluation is rather complicated,
since it includes summation over Matsubara frequencies, integration over the
momentum $k$ and the averaging over the angles. Here we will discuss two
limiting cases in which Eq. (\ref{crit_general}) can be simplified.

a) $h\tau \ll 1$ (dirty case). This limit corresponds to a ferromagnet with
a weak exchange field $h$ or to an alloy like that used in Ref. \cite{Ryaz},
for which the condition $h\geq T_{c}$ is satisfied. In this case the
condition $h\tau \ll 1$ implies that the quantities $\Delta \tau $ and $%
T\tau $ are also small. From  Eq. (\ref{B}) one can determine the coefficient $B$; in this limit $B\cong \langle \mu \gamma \rangle l/\left[
(l^{2}/3)(k^{2}+k_{+}^{2})\right] $, where $\kappa _{+}^{2}=2(\omega -ih)/D$
and $D=v_{F}l/3$ is the diffusion coefficient. Using Eq. (\ref{crit_general}%
) we obtain for the normalized critical current $J_{c}$  
\begin{equation}
J_{c}=\frac{3}{4}\langle \mu \gamma \rangle ^{2}{\rm Re}(2\pi T)\sum_{\omega
>0}{\frac{1}{\kappa _{+}l\sinh (\kappa _{+}d)}\frac{\Delta ^{2}}{\Delta
^{2}+\omega _{m}^{2}}}\;.  \label{crit_dirty}
\end{equation}

This is the usual expression for the critical current obtained from the
Usadel equation in the case of a weak proximity effect ({\it cf.} Refs. \cite
{Buzdin1,Ryaz,Schon}). In Fig.2 and Fig.3 we
plot $J_{c}$ as a function of $T$ and $d$. As has been shown in the previous
studies, the function $J_{c}$ is a rapidly decaying with $T$ and $d$
function which undergoes several oscillations. This oscillatory behavior of
the critical current may explain the change from a $0$-phase state to a $\pi 
$-phase state observed in Ref. \cite{Ryaz}. 


We consider now another interesting case.

b) $h\tau\gg 1$

This condition corresponds to most of the experiments performed on S/F
systems, in which F is a ''strong'' ferromagnet like $Ni$ or $Fe$. It does
not necessarily  mean that we are analyzing the clean case ({\it i.e} $%
T_{c}\tau >1$), since the value of the exchange field $h$ can be much larger
than $T_{c}$. For example, if the mean free path $l$ equals $\sim 300\AA $,
then $\tau ^{-1}\sim 300$K$\gg T_{c}$, whereas $h\tau \geq 1$ (we take $%
v_{F}=2.10^{7}$cm/s). Therefore in the limit $h\tau \gg 1$ we can deal in
principle with an arbitrary value of $\tau T_{c}$ (although realistic
materials and samples correspond to the case $T_{c}\tau \ll 1$). It is worth
mentioning that in this case the use of the Usadel equation is not
justified. 

The condition $h\tau \gg 1$ implies that $\kappa _{\omega }\gg 1$, $\kappa
_{\omega }/M\ll 1$ and, as one can see from Eq. (\ref{B}), $B\cong \kappa
_{\omega }(\gamma \mu l)/M$. Performing the integration over $k$, we find
for $J_{c}$

\begin{equation}  \label{crit_clean}
J_c=\frac{1}{4}(2\pi T){\rm Re}\sum_{\omega>0}{f_s^2\left\langle\frac{%
\mu\gamma^2}{\sinh(\kappa_\omega d/\mu l)}\right\rangle}\; ,
\end{equation}

One can see from Eq. (\ref{crit_clean}) that the critical current oscillates
with varying $d$ or $h$ ({\it cf.} Ref. \cite{Buzdin} where the case $d\ll l$
was considered) and decays with increasing $d$ over the mean free path $l$.
In Fig.4 and Fig.5 we plot the dependence of 
$J_{c}$ on $d$ and $T$ calculated numerically from Eq. (\ref{crit_clean}).
We see that in this limit the critical current does not oscillate with the
temperature if the exchange field is temperature independent ( this
assumption is quite reasonable in the case of transition metals as $Fe$ or $%
Ni$). It can change sign in a hypothetical case $T_{c}\tau >1$. In the limit 
$d/l\gg 1$ the critical current is exponentially small and one can perform
the angle averaging in Eq. (\ref{crit_clean}). Thus, we find in this limit 
\begin{equation}
J_{c}=(\pi T)\sum_{\omega >0}{f_{s}^{2}\gamma ^{2}(1)\frac{\sin (2hd/v_{F})}{%
(2hd/v_{F})}\exp {\left( -(d/l)(1+2\omega \tau )\right) }}\;.
\label{crit_clean_lim}
\end{equation}
It follows from Eq. (\ref{crit_clean_lim}) that 
the critical current oscillates as a function of $d$ or $h$ and decays with $%
d$ exponentially if $d>l$ and as $(hd/v_{F})^{-1}$ if $d\cong l$. This power law dependence of $I_{c}$ on $h$ can be even weaker
if $\gamma $ depends on $\mu $ ($\gamma $ has a maximum at $\mu =1$ and
decays with decreasing $\mu $).



\section{S/F/S-junction with an nonhomogeneous magnetization}

\label{section_sfs} In this section we consider a nonhomogeneous
magnetization in the magnetic region. This non-homogeneity can be due to
domain walls or, as in the case of experiments on giant magnetic resistance
(GMR), due to an artificial layered magnetic structure. In principle,
variation of the magnetic moment on the coordinates can be rather
complicated, which makes explicit calculations difficult. To simplify the
consideration we restrict ourselves with the cases of a magnetic spiral
structure in the F region with a wave vector $Q$ and calculate the
dependence of the critical Josephson current $I_{c}$ in a S/F/S junction on
the wave vector $Q$. 

Below we consider only the dirty case ($h\tau <1$) and assume again a weak
proximity effect. This means that $\gamma $ must be small enough: $\gamma
\ll \sqrt{h\tau }$. In the limit of small $h\tau $ one can use the Usadel
equation for finding the condensate function $\hat{f}$. However in the case
of a rotating magnetization (or equivalently a rotating exchange field $h$)
we need to generalize our approach because not only singlet correlators as $%
\langle \psi _{\uparrow }\psi _{\downarrow }\rangle $ are induced in the F
region, but also correlators of the type $\langle \psi _{\uparrow }\psi
_{\uparrow }\rangle $ become nonzero (triplet component). In this case we
introduce new $4\times 4$ matrices for the quasiclassical Green's functions
(see Appendix). The $4\times 4$ condensate Green's function $\check{s}$ (to
be more exact, its symmetrical part) obey the generalized Usadel equation in
the Matsubara representation 
\begin{equation}
-iD\check{g}_{0}.\partial _{xx}^{2}\check{s}+\left[ \check{M}_{h},\check{s}%
\right] =0\;,  \label{usadel_sfs}
\end{equation}
where $\check{g}_{0}=\hat{\tau}_{3}\otimes \hat{\sigma}_{0}$, $D=v_{F}l/3$
is the diffusion coefficient, and $\check{M}_{h}=\hat{\tau}_{3}\otimes (\hat{%
\sigma}_{0}i|\omega _{m}|+\hat{\sigma}_{3}h{\rm sgn}\omega _{m}\cos \alpha )-%
\hat{\tau}_{0}\otimes \hat{\sigma}_{2}h{\rm sgn}\omega _{n}\sin \alpha $.
Here $\alpha =Qx$ is the angle between the $z$-axis and the direction of $h$%
. As in the previous section, we neglect corrections to the ''normal''
Green's function $\check{g}_{0}$ due to the proximity effect. 

Eq. (\ref{usadel_sfs}) is a linear matrix differential equation with space
dependent coefficients. This spatial dependence can be excluded from the
consideration by making a rotation in the Nambu-spin space and by
introducing a new matrix $\check{s}_n$: $\check{s}=\check{U}\check{s}_{n}\check{U}^{+}$,
where $\check{U}=\hat{\tau}_{0}\otimes \hat{\sigma}_{0}\cos (\alpha /2)+i%
\hat{\tau}_{3}\otimes \hat{\sigma}_{1}\sin (\alpha /2)$. After the rotation
Eq. (\ref{usadel_sfs}) acquires the form 
\begin{equation}
\partial _{xx}^{2}\check{s}_{n}-(Q^{2}/2)\left( \check{s}_{n}-\check{A}%
\check{s}_{n}\check{A}^{+}\right) +Q\left( \check{A}\partial _{x}\check{s}%
_{n}+\partial _{x}\check{s}_{n}\check{A}^{+}\right) -\left[ \hat{\tau}%
_{3}\otimes \left( \hat{\sigma}_{0}|\omega _{m}|-ih{\rm sgn}\omega _{m}\hat{%
\sigma}_{3}\right) ,\check{s}_{n}\right] _{+}=0\,,  \label{rot_usadel}
\end{equation}
where $\check{A}=i\hat{\tau}_{3}\otimes \hat{\sigma}_{1}$ and the last term
is an anticommutator. If $Q=0$, Eq. (\ref{rot_usadel}) coincides with Eq. (%
\ref{usadel_sfs}) and can be easily solved. The critical current $I_{c}$ has
been calculated in the previous section, and its dependence on $T$ and $d$
is presented in Fig.2 and Fig.3. It is seen
that the critical current $I_{c}$ changes the sign at some $h$ of the order
of the Thouless energy, $\epsilon _{d}=D/d^{2}$. The characteristic value of 
$h$ for the transition to the $\pi $-state increases with increasing $T$.
This result is well known for both types of Josephson (equilibrium and
nonequilibrium) junctions in which the sign-reversal effect of $I_{c}$ takes
place. Earlier than in S/F/S junctions, this effect was observed in
4-terminal S/N/S Josephson junctions where a voltage $V$ was applied between
the normal reservoirs and therefore an additional dissipative current flows
between the N reservoirs \cite{Wees} were possible. 

The critical current changes sign in these junctions due to a shift of the
distribution function in the N electrode with respect to the distribution
function in the superconductors. In contrast to this case, the sign reversal
effect in S/F/S junctions is realized at equilibrium conditions. However
there is a formal analogy between these two cases because the formulae for
the critical current can be reduced to each other by shifting the energy
scale and by replacing $h$ $\rightarrow $ $eV$ (this analogy was noted in
Refs.\cite{VP,Schon}).

In the case of a finite $Q$, the formulae for $\check{s}$ become more
complicated. In order to make them more transparent, we assume that the
overlap of the condensate functions $\check{s}$ induced by the different
superconductors is weak. This means that $\max \{h,T\}$ should be greater
than $\epsilon _{d}.$ Then, we represent the solution of Eq. (\ref
{rot_usadel}) for $\check{s}$ \ as a sum of two functions

\begin{equation}
\check{s}(x)=\check{U}\left[ \check{S}.\check{s}_{n}(d/2+x).\check{S}^{+}+%
\check{S}^{+}.\check{s}_{n}(d/2-x).\check{S}\right] \check{U}^{+}\;.
\label{solution_for_f}
\end{equation}
The matrix $\check{S}$ allows one to take into account the phase difference $%
\varphi $ between the superconductors: $\check{S}=(\hat{\tau}_{0}\cos
(\varphi /4)+i\hat{\tau}_{3}\sin (\varphi /4))\otimes $ $\hat{\sigma}_{0}$.
The first and second term in Eq. (\ref{solution_for_f}) come from the
superconductor at $x=-(d/2)$ and $x=+(d/2)$ respectively. The function $%
\check{s}_{n}(x)$ is a solution of Eq. (\ref{rot_usadel}) for an infinite
S/F system with a vanishing phase $\varphi =0$. The boundary condition for
the new matrix $\check{s}_{n}$ has the form 
\begin{equation}
\partial _{x}\check{s}_{n}+(Q/2)\left[ \check{A}\check{s}_{n}+\check{s}_{n}%
\check{A}^{+}\right] =-(\tilde{\gamma}/l)\check{f}_{s}\;.
\label{new_boundary}
\end{equation}
Here $\check{f}_{s}=i\check{\tau}_{2}\otimes \check{\sigma}_{3}f_{s}$ and $%
\tilde{\gamma}=3\langle \gamma \mu \rangle $; $f_{s}$ and $\gamma $ have
been defined in the previous section. It is not difficult to see that the
function $\check{s}_{n}(x)$ has the following form

\begin{equation}
\check{s}_{n}(x)=i\hat{\tau}_{2}\otimes (\hat{\sigma}_{0}S_{0}(x)+\hat{\sigma%
}_{3}S_{3}(x))+i\hat{\tau}_{1}\otimes \hat{\sigma}_{1}T(x)\; ,
\label{solution for f_o}
\end{equation}
where the functions $S_{0}(x),S_{3}(x)$ and $T(x)$ are the amplitudes of the
singlet and triplet component, respectively. All these functions may be
represented as a sum of three exponentials corresponding to the eigenvalues
of Eq.(\ref{rot_usadel}).\ For example, the expression for $S_{3}(x)$ is

\begin{equation}
S_{3}(x)=S_{3+}\exp (-\kappa _{+}x)+S_{3-}\exp (-\kappa _{-}x)+S_{3l}\exp
(-\kappa _{l}x)\;.  \label{solution for S}
\end{equation}
Identical formulae may be written for the functions $S_{0}(x)$ and $T(x)$
with the factors in front of the exponentials denoted as $S_{0\pm },S_{0l}$
and $T_{\pm },T_{l}$ correspondingly. Analytical expressions for the
coefficients and eigenvalues can be obtained in the limits of small and
large $Q$, {\it \ i.e.} $DQ^{2}\ll h$ or $DQ^{2}\gg h$. In the limit of
small $Q,$ we find after some algebra 
\begin{equation}
\kappa _{\pm }^{2}=2(|\omega _{m}|\mp {\rm sgn}\omega ih)/D\;,\;\;\kappa
_{l}^{2}=2|\omega _{m}|/D+Q^{2}  \label{eigenvalues}
\end{equation}
and 
\begin{equation}
S_{3\pm }=\mp S_{0\pm }=\tilde{\gamma}f_{s}/2(\kappa _{\pm }l)\;\;,T_{l}=%
\frac{1}{2}(\tilde{\gamma}f_{s})\frac{(\kappa _{-}-\kappa _{+})Q}{l(\kappa
_{+}\kappa _{-})\kappa _{l}}\;,\;\;S_{3l}=(\tilde{\gamma}f_{s})\frac{(\kappa
_{+}-\kappa _{-})Q^{2}}{l(\kappa _{+}\kappa _{-})\kappa _{h}^{2}}{\rm sgn}\omega\;,
\label{SandT}
\end{equation}
where $\kappa _{h}^{2}=-2ih/D$. 

We note some new important features that appear at finite $Q.$ If $Q$ is
zero, only the first two terms in Eq. (\ref{solution for S}) are nonzero and
the decay is characterized by a short length $\xi _{F}=\sqrt{D/h}$ (in case
of large enough $h$). If $Q$ is finite, an additional term (the last term)
appears in the formula for $S_{3}(x)$ which, at low temperatures, decays
over a much larger length of the order $\sqrt{D/2\pi T}.$ Alongside with the
last term, the triplet component becomes nonzero. The triplet component
contains also a long-range term $T_{l}\exp (-\kappa _{l}x)$ which increases
with increasing $Q$. Due to the last term in Eq. (\ref{solution for S}), the
dependence of the critical current on $h$ (or $T$) is drastically modified
even at small $Q$. 

In order to calculate the current $I_{J}$, we can use the general expression
Eq. (\ref{current_sfs}). In the dirty limit it can be written as follows 
\begin{equation}
I_{J}=({\cal S}/16\rho )(2\pi i)(2T){\rm Tr}\left( \hat{\tau}_{3}\hat{\sigma}%
_{0}\right) \sum_{\omega }{\check{s}\partial _{x}\check{s}}=({\cal S}/16\rho
)(2\pi i)(2T){\rm Tr}\left( \hat{\tau}_{3}\hat{\sigma}_{0}\right) (\tilde{%
\gamma}/l)\sum_{\omega }{\check{s}\check{f_{s}}|_{x=d/2}}\;.
\label{dirty_current}
\end{equation}
One can show that this expression does not change its form under the $\check{%
U}$-transformation and $\check{s}$ in Eq. (\ref{dirty_current}) may be
replaced by $\check{s}_{n}(d/2)$. Performing simple operations, we obtain
for the current an expression with the same form as Eq. (\ref{jos_current})
with 
\begin{equation}
I_{c}=({\cal S}/\rho l)\tilde{\gamma}^{2}\sum_{\omega >0}f_{s}^{2}\left[ 
\frac{\exp (-\kappa _{+}d)}{\kappa _{+}l}+\frac{(Ql)^{2}}{2(3h\tau )^{3/2}}%
\exp (-\kappa _{l}d)\right] \;.  \label{dirty_crit_current}
\end{equation}
The first term in the brackets corresponds to the term $\left[ 2(\kappa
_{+}l)(\sinh \kappa _{+}d)\right] ^{-1}$ in Eq. (\ref{crit_general}) in the
limit of a large exponent. It decays with increasing $d$ over the short
characteristic length $\xi _{F}=\sqrt{D/h}$. The second term in Eq. (\ref
{dirty_crit_current}) is caused by the rotation of $h$ along the $x$-axis.
It decays with $d$ over the characteristic length $\kappa _{l}^{-1}$, which
can be much longer than $\xi _{F}$. Therefore this term leads to a drastic
change of the critical current. We calculated numerically the critical
current $I_{c}$ and presented its temperature dependence for $h\tau =0.06$
in Fig.6. One can see that for this choice of $h$ the
critical current is negative at $Q=0$ ($\pi $-junction). However it becomes
positive at some finite $Ql$ smaller than $h\tau $. With increasing $Q$ the
long-range term in Eq. (\ref{solution for S}) and the triplet component
increase, reach a maximum and then decrease again to zero at large enough $Q$%
.



In the limit of large $Q$, the coefficients $S_{3\pm}$, $S_{0\pm}$ and the
triplet components are small. The coefficient $S_{3l}=\tilde{\gamma}f_s/(kl)$
has the same form as in the absence of $h$. For the eigenvalues we obtain

\[
\begin{array}{cc}
\kappa _{\pm }=\pm iQ+\sqrt{2\omega_{m}/D}\;\; , & \kappa_l =\sqrt{%
2(\omega_{m} /D)+4h^{2}/(DQ)^{2}}.
\end{array}
\]
We see that in the limit of large $Q$ the solution for $\check{f}$ has the
same form as in a S/N/S structure (no exchange field), i.e. the term $\left(
sf_{S}/\theta \right) \exp (-\kappa_l x)$ dominates. The first two terms in
the singlet component $S_{3}(x)$ which contribute to the Josephson current (
see Eq. (\ref{solution for S})) are small. They oscillate rapidly in space
and decay over a large distance of the order of $\xi _{T}=\sqrt{D/2\pi T}$
(in the limit $DQ^{2}<h$). In the main approximation in the parameter $%
(h/DQ^{2})$ the temperature dependence of the critical current $I_{c}$ is
the same as for a S/N/S junction and we do not present this dependence here.

\section{The S/F-I-F/S System}

\label{section_sfisf}

In this section we consider a layered system consisting of two F/S bilayers
separated by an insulating layer (see Fig.1). In this case the Josephson
critical current is determined by the transparency of the insulating layer
and depends on the relative orientation of magnetization in the F layers.


We assume that the F and the S layers $d_{F,S}$ are thin enough: $%
d_{F,S}<\xi _{F,S}$, where $\xi _{F}=\sqrt{D/h}$ and $\xi _{S}=\sqrt{%
D/\Delta }$. First, we analyze the case of a high S/F interface
transparency, i.e. $R_{S/F}<\rho _{F}/\xi _{F}$. Under these conditions all
the Green's functions are nearly constant in space and continuous across the
S/F interface. 

In order to find the Green's functions $\check{g}^{R(A)}$, we multiply the
components (1,1) and (2,2) of the matrix equation (\ref{usadel}) (the Usadel
equations) by the density of states $\nu _{F,S}$ in the F and the S layers
respectively, and integrate over the thickness of the bilayers. Neglecting
the influence of one bilayer on the other (this means that $\left( \check{g}%
\partial _{x}\check{g}\right) =0$ at the F/I interface), we obtain the
following equation:

\begin{equation}
{\rm sgn}\omega \left[ \check{M}_{h},\check{{\bf g}}\right] +\left[ \hat{%
\Delta}_{S}\otimes \hat{\sigma}_{3},\check{{\bf g}}\right] =0\,\,,
\label{usa1}
\end{equation}
Here the matrix $\check{M}_{h}$ has the same structure as in Eq. (\ref
{usadel_sfs}), but $h$ has been replaced by $h_{F}=h\left( \nu
_{F}d_{F}\right) /\left( \nu _{F}d_{F}+\nu _{S}d_{S}\right) $; $\hat{\Delta}%
_{S}=\hat{\Delta}\left( \nu _{S}d_{S}\right) /\left( \nu _{S}d_{S}+\nu
_{F}d_{F}\right) $. We assume that the vector ${\bf h}$ in the left layer is
oriented along the z-axis and has the components $h\left( 0,\sin \alpha
,\cos \alpha \right) $ in the right electrode. One can simplify Eq. (\ref
{usa1}) in the right bilayer with the help of the transformation (\ref
{rotation}). In this case one obtains for the both layers the same equation: 
\begin{equation}
\left[ \hat{\tau}_{3}\otimes (\epsilon \hat{\sigma}_{0}+h_{F}\hat{\sigma}%
_{3}),\check{g}\right] +\left[ \hat{\Delta}_{S}\otimes \hat{\sigma}_{3},%
\check{g}\right] =0\;.  \label{usadelrot1}
\end{equation}
We can solve Eq.(\ref{usadelrot1}) by making the ansatz 
\begin{equation}
\check{g}=\hat{\tau}_{3}\otimes (a_{0}\hat{\sigma}_{0}+a_{3}\hat{\sigma}%
_{3})+\hat{\Delta}_{S}\otimes (b_{0}\hat{\sigma}_{0}+b_{3}\hat{\sigma}%
_{3})\,.  \label{ansatz1}
\end{equation}
From Eq. (\ref{usadelrot1}) and the normalization condition (\ref
{normalization}) one can obtain the coefficients $a$'s and $b$'s. In the
left bilayer $\check{g}$ is given by the expression (\ref{ansatz1}) while in
the right bilayer it is given by $\check{g}^{(r)}=\check{U}^{+}\widetilde{%
\check{g}}^{(r)}\check{U}$, i.e. 
\[
\check{g}^{(r)}=\hat{\tau}_{3}\otimes (a_{0}\hat{\sigma}_{0}+a_{3}\cos
\theta \hat{\sigma}_{3})-\hat{\tau}_{0}\otimes a_{3}\sin \theta \hat{\sigma}%
_{2}+\hat{\Delta}_{S}\otimes (b_{0}\cos \theta \hat{\sigma}_{0}+b_{3}\hat{%
\sigma}_{3})-\hat{\tau}_{3}\hat{\Delta}_{S}\otimes ib_{0}\sin \theta \hat{%
\sigma}_{1}\,.
\]
According to Eq. (\ref{current}) only the coefficients $b_{0}$ and $b_{3}$
will enter in the expression for the Josephson current, and they are given
by 
\[
(b_{3})_{l,r}=\frac{1}{2}\left( \frac{1}{\xi _{+}}+\frac{1}{\xi _{-}}\right)
_{l,r}\;{\rm and}\;(b_{0})_{l,r}=\frac{1}{2}\left( \frac{1}{\xi _{+}}-\frac{1}{%
\xi _{-}}\right) _{l,r}\;,
\]
where $\xi _{\pm }=\sqrt{\epsilon _{\pm }^{2}-|\Delta _{S}|^{2}}$, and $%
\epsilon _{\pm }=i\omega _{m}\pm h$. By writing $\Delta _{S}=|\Delta
_{S}|\exp (i\varphi )$ in the right side one obtains the following
expression for the critical current 
\begin{equation}
eV_{c}(\alpha )\equiv eI_{c}R_{b}=2\pi T\Delta _{l}\Delta
_{r}\sum_{m>0}\left\{ {\rm Re}\left( \frac{1}{\xi _{m}}\right) _{l}{\rm Re}%
\left( \frac{1}{\xi _{m}}\right) _{r}-\cos \theta {\rm Im}\left( \frac{1}{%
\xi _{m}}\right) _{l}{\rm Im}\left( \frac{1}{\xi _{m}}\right) _{r}\right\} \;,
\label{crit_current_sfisf}
\end{equation}
where $\xi _{n}=\sqrt{\left( \omega _{m}+ih_{F}\right) ^{2}+\Delta _{S}^{2}}$
and $R_{b}$ is the tunnel resistance of the I layer. Formula (\ref
{crit_current_sfisf}) coincides with the formula presented in Ref. \cite
{kulic}. In the latter work the authors considered a Josephson junction
consisting of two magnetic superconductors with an oscillating magnetic
order. Thus, we have shown that the system of Fig.7 and that
of Ref. \cite{kulic} are equivalent. However the authors of Ref. \cite{kulic}
did not consider some interesting properties of such structures. We note
that the same structure was also analyzed in Ref.\cite{Koshina}, where the
critical current was calculated for different S/F interface transparencies.
The authors have found the conditions under which the system undergoes a
transition to the $\pi $ state; however they analyzed only the case of
parallel magnetization. 

Here we consider two limiting cases: a) a parallel relative orientation of
the magnetizations, i.e. $\alpha =0$ and b) an antiparallel orientation: $%
\alpha =\pi $.

In the case $\alpha =0$ according to Eq. (\ref{crit_current_sfisf}), the
critical current is given by the expression 
\begin{equation}
eV_{c\uparrow \uparrow }\equiv eI_{c\uparrow \uparrow }R_{b}=4\pi T\Delta
_{S}^{2}\sum_{m}\frac{\omega _{m}^{2}+\Delta _{S}^{2}-h_{F}^{2}}{\left(
\omega _{m}^{2}+\Delta _{S}^{2}-h_{F}^{2}\right) ^{2}+4\omega
_{m}^{2}h_{F}^{2}}\;.  \label{crit_curr_para}
\end{equation}
In writing Eq. (\ref{crit_curr_para}) we assumed that $h_{F}$ and $|\Delta
_{S}|$ are the same in both bilayers (symmetric structure). The dependence
of the critical current on the exchange field $h_{F}$ was presented in Ref. 
\cite{BVE}. At $T=0$ the current $I_{c}$ is constant up to the value $%
h_{F}=\Delta _{0}$ where it drops to zero; $\Delta _{0}$ is the effective
energy gap $\Delta _{S}$ at zero temperature and zero exchange field. This
is a consequence of the fact that the order parameter $\Delta $ is also
constant. We do not consider here a possible transition to the LOFF phase
predicted by Larkin and Ovchinnikov (LO) \cite{LOv} and Fulde and Ferrell
(FF) \cite{Fulde} for the region $0.755\Delta _{S0}<h_{F}$. We argue that
since the homogeneous superconducting state in this region is a metastable
state, its realization is possible. Nevertheless our result is definitely
valid for the region of small $h_{F}$, and a possible transition to the LOFF
would manifest itself in a drop of the the critical current.

More interesting is the case when the relative orientation of the
magnetizations is antiparallel, i.e. $\alpha =\pi .$ Then, the critical
current is given by the expression 
\begin{equation}
eV_{c\uparrow \downarrow }(\pi )\equiv eI_{c\uparrow \downarrow }R_{b}=4\pi
T\Delta _{S}^{2}\sum_{m}\frac{1}{\sqrt{\left( \omega _{m}^{2}+\Delta
_{S}^{2}-h_{F}^{2}\right) ^{2}+4\omega _{m}^{2}h_{F}^{2}}}\;.
\label{crit_curr_anti}
\end{equation}
In this case the dependence of $I_{c}$ on $h_{F}$ is completely different
from that given by Eq. (\ref{crit_curr_para}). The critical current
determined by Eq. (\ref{crit_curr_anti}) increases with increasing $h_{F}$
(i.e. with increasing either $h$ or $d_{F}$) and even diverges at zero
temperature when $h_{F}\rightarrow \Delta _{S}$. Of course, there is no real
divergence of $I_{c}$ since, for example, finite temperatures  smear out
this divergency. The dependence of $eV_{c}/\Delta _{0}$ on $h$ was presented
in Ref.\cite{BVE}. The critical current has a maximum at some value of $h_{F}
$ close to $\Delta _{0}$. With decreasing $T$ the maximum value of $I_{c}$
increases and its position is shifted towards $\Delta _{0}$. For arbitrary
relative orientations of magnetizations the expression for $V_{c}(\alpha )$
can be presented in the form 
\begin{equation}
V_{c}(\alpha )=V_{c\uparrow \uparrow }\cos ^{2}(\alpha /2)+V_{c\uparrow
\downarrow }\sin ^{2}(\alpha /2)\;.  \label{generalalpha}
\end{equation}
Therefore, the singular part is always present and its contribution reaches
100\% at $\alpha =\pi $. 
All the conclusions given above remain valid also for two magnetic
superconductors with uniformly oriented magnetizations in each layer. We
note that in contrast to the case of the spiral structure analyzed in Ref. 
\cite{kulic}, no $\pi $-state appears in our model for any effective
exchange field $h_{F}\leq \Delta _{0}$ (at larger $h_{F}$ the
superconductivity is destroyed). As in the previous case of parallel
orientations, the state with $h_{F}=\Delta _{0}$ might be unreachable for
the antiparallel orientation due to the appearance of the inhomogeneous LOFF
state. However the singular behavior of $I_{c}$ can be realized at smaller
values of $h$ in a
structure with large enough S/F interface resistance $R_{S/F}$. In this case
the bulk properties of the S film are not changed by the proximity of the F
film (to be more precise the condition $R_{S/F}>(\nu_Fd_F/\nu_Sd_S)\rho_{F}\xi_{F}$ must be satisfied; $\rho_F$ is the specific resistance of the F film). Then, as one can readily show \cite{mcmillan}, a subgap $\epsilon
_{sg}=\left( D\rho \right) _{F}/\left( R_{S/F}d_{F}\right) $ arises in the F
layer. The Green's functions in the F layer have the same form as in Eq. (%
\ref{ansatz1}) with $\Delta _{S}$ replaced by $\epsilon _{sg}$. The
singularity in $I_{c}(h_{F})$ occurs at $h_{F}$ equals to $\epsilon _{sg}$,
and the LOFF state does not arise because the subgap $\epsilon _{sg}$ is not
determined by the self-consistency equation.

For completeness we note that the effect of the relative orientation of
magnetization in the F films on the critical temperature $T_c$ of the
superconductor was analyzed in Refs.\cite{Tagirov,Buzdin3} for a F/S/F
structure.

\section{Orbital effects}

In the preceding sections we have presented the formulae for the condensate
function $\hat{s}$ (or $\check{s}$) in the ferromagnetic regions induced by
the proximity effect. The amplitude of $\hat{s}$ is determined by the
interface transparency, {\it i.e.} by the parameter $\gamma $, and the
penetration length depends essentially on the parameter $h\tau $. The
internal magnetic field $B$ of the ferromagnet, induces screening currents
and leads to some suppression of the condensate function. We have neglected
this suppression due to Meissner currents (orbital effects). In order to
understand why this approximation can be justified we estimate now the
magnitude of these effects.

In the dirty limit the depairing rate due to Meissner current is determined
by the energy $Dp_{s}^{2}$, where $p_{s}=A\left( d\right) /\phi _{0}$ is the
condensate momentum, $A\left( x\right) =Bx$ is the vector potential and $%
\phi _{0}$ is the magnetic flux quantum. The depairing factor can be
neglected in the Usadel equation provided the condition 
\begin{equation}
Dp_{s}^{2}\ll h  \label{cond_dep}
\end{equation}
is satisfied. For example, for $B=1kG$, $d=100$\AA , $v_{F}=2.10^{7}$cm/s
and $l\sim d$ we obtain $Dp_{s}^{2}\sim 50$mK. If the condition Eq. (\ref
{cond_dep}) is met, the condensate function $\hat{s}(x)$ (to be more exact,
its Fourier transform) in the ferromagnet is given by Eqs. (\ref{fourier_s}-%
\ref{M}). 

Due to the condensate penetration and the intrinsic magnetic field of the
ferromagnet, the Meissner currents arise in the F region. In order to
analyze this issue in more detail, we consider the S/F system of Fig.8.


If we consider the diffusive regime, the condensate function can be found as
it was done in section II or directly from the Usadel equation. In this
limit ($h\tau \ll 1$) it has the form: 
\begin{equation}
s_{\omega }(x)=3\langle \mu \gamma \rangle f_{s}\frac{\cosh \kappa _{+}(d-x)%
}{(\kappa _{+}l)\sinh (\kappa _{+}d)}\;.  \label{s_5}
\end{equation}
The current density is expressed in terms of $s(x)$ as 
\begin{equation}
j(x)=-\sigma (Bx/\phi _{0})(2\pi T){\rm Re}\sum_{\omega >0}s_{\omega
}^{2}(x)\;.  \label{curr_den}
\end{equation}


In Fig.9 we plot the spatial dependence of the current density $%
j(x)$ which is spontaneously induced in the ferromagnetic film. We can see
that $j(x)$ changes sign with varying $x$. According to the results of
section II, in the case $h\tau >1$, the current density changes sign many
times on the mean free path.

Integrating the current density $j(x)$ given by Eq. (\ref{curr_den}), we
find the total current $I=\int_{0}^{d}{\rm d}xj(x)$

\begin{equation}
I=-\frac{1}{4}\left( 3\langle \mu \gamma \rangle \right) ^{2}\sigma
(Bd^{2}/\phi _{0})(2\pi T){\rm Re}\sum_{\omega >0}{\frac{f_{s}^2}{(\kappa
_{+}l)^{2}\sinh ^{2}(\kappa _{x}l)}\left[ \frac{\sinh ^{2}(\kappa _{+}d)}{%
(k_{+}d)^{2}}+1\right] }\;.  \label{total_curr}
\end{equation}
In Fig.10 and Fig.11 we presented the total
current $I$ as a function of $d$ and $h\tau $ respectively. It is seen that
the total current also changes sign, {\it i.e.} in the ferromagnetic film
either a diamagnetic or paramagnetic current is induced depending on the
relation between $d$ and $\xi _{h}$.


In the analysis presented here it was assumed that the exchange field $h$ is
homogeneous. In a multidomain ferromagnet one expects a more complicated
spatial distribution of the Meissner current.

\section{Conclusion}

We analyzed specific features of a supercurrent in
superconductor/ferromagnet structures. In section II we have calculated the
Josephson current $I_{J}$ in a S/F/S junction. It turns out that the product 
$h\tau $ of the splitting energy $h$ and the momentum relaxation time $\tau $
is an important parameter, which determines the approach to be used in the
problem. In the dirty limit, i.e. when $h\tau \ll 1$,  $I_{J}$ can be
obtained from the Usadel equation (Refs. \cite{Ryaz,Kupr,Schon}). In this
limit, the change of sign of the critical current $I_{c}$ occurs if the
thickness of the F layer $d$ is of the order of $\xi _{F}=\sqrt{D/h}$. The
condensate function in the F layer decays exponentially over the length $\xi
_{F}$ and undergoes oscillations with the same period. In the opposite limit
($h\tau \gg 1$) the condensate function oscillates in space with the period $%
v_{F}/h$ (as in the pure ballistic case considered in Ref.\cite{Buzdin}) and
decays exponentially on the mean free path $l$. The critical current $I_{c}$
decreases with $h$ as a power-law function and is not exponentially small if 
$d\sim l$.  

We have also studied the influence of different inhomogeneous magnetic
structures on the critical current through S/F structures. In section III we
considered a S/F/S sandwich with an inhomogeneous magnetic order in the F
layer described by a vector $Q$. In the case $Q=0$ we obtained the
well-known transition from the $0$-phase to the $\pi $-phase state. We have
also shown that even for small values of $Q$ and not too low temperatures
this transition may not take place (Fig.6). The reason for
a qualitative change of the $I_{c}(h_{F})$ is a long-range term in the
singlet component of the condensate function $\check{f}$. This term arises
together with the triplet component if $Q$ differs from zero. The long-range
part of $\check{f}$ decays in the F film on a length of the order $\sqrt{%
D/2\pi T}$, which can be much longer than the characteristic length ($\sim 
\sqrt{D/h}$) of the decay of $\check{f}$ in a homogeneous F layer ($Q=0$).

Our results may be applied to ferromagnets containing domain walls and
magnetic multilayers with nonhomogeneous magnetic structures. We used the
quasiclassical Green's function approach to describe such
structures in a  quantitative  way. In section IV it was shown that for an antiferromagnetic
configuration in the S/F-I-S/F junction, the dependence $I_{c}(h)$ shows an
anomalous behavior: the critical current increases with increasing $h$ or $%
d_{F}$. This means that the Josephson critical current in a junction formed
by two ferromagnet-superconductor bilayers may be even larger than the
critical current in a similar Josephson junction S/I/S. In the last section
we have considered a S/F bilayer structure. The Meissner currents which are
spontaneously excited due to the internal field of the ferromagnetic film
have been calculated. The  current density oscillates along the $x$-axis and
the total Meissner current in the F film may be either diamagnetic or
paramagnetic depending on the thickness $d$ and on the exchange field $h$.
All the effects analyzed in our work can be verified experimentally.

\section*{Acknowledgments}

. We thank SFB 491 {\it Magnetische Heterostrukturen} for financial support.

\section*{Appendix}

In this part we present some general formulae for superconductivity in the
presence of an exchange field. We consider structures in which the
superconducting pairing and the exchange interaction of electrons with
ordered, localized magnetic moments take place. The Hamiltonian describing
the system under consideration has the form:

\begin{equation}
\hat{H}=\sum_{\{p,s\}}\left\{ a_{sp}^{+}\left[ \left( \left( \xi _{p}\delta
_{pp^{\prime }}+eV\right) +U_{imp}\right) \delta _{ss^{\prime }}-\left( {\bf %
h}.\sigma \right) \right] a_{s^{\prime }p^{\prime }}-\left( \Delta a_{%
\overline{s}\overline{p}}^{+}a_{s^{\prime }p^{\prime }}^++c.c.\right) \right\} 
\end{equation}
Here the summation is carried out over all momenta $(p,p^{\prime })$ and
spins $(s,s^{\prime })$, $\xi _{p}=p^{2}/2m-\epsilon _{F}$ is the kinetic
energy counted from the Fermi energy $\epsilon _{F}$, $V$ is a smoothly
varying electric potential, $U_{imp}=U(p-p^{\prime })$ is a potential
describing the interaction of electrons with non-magnetic impurities, $h$ is
an effective ``magnetic field'' caused by the exchange interaction of spins
of the free electrons with spins of the localized magnetic moments. The
notation $\overline{s}$, $\overline{p}$ means inversion of both spin and
momentum. The order parameter $\Delta $ must be determined
self-consistently. In order to define the Green's function in a customary
way we introduce new operators $c_{ns}^{+}$ and $c_{ns}$, which are related
to the creation and anhilation operators $a_{s}^{+}$ and $a_{s}$ by the
relation (we drop the index $p$ related to the momentum) 
\[
c_{ns}=\left\{ 
\begin{array}{c}
a_{s},n=1 \\ 
a_{\overline{s}}^{+},n=2\;.
\end{array}
\right. 
\]
The index $n$ operates in the particle-hole (Nambu) space, while the index $s
$ operates in the spin space. The operators $c_{ns}$ obey the commutation
relations 
\begin{eqnarray}
c_{ns}c_{n^{\prime }s^{\prime }}^{+}+c_{n^{\prime }s^{\prime }}^{+}c_{ns}
&=&\delta _{nn^{\prime }}\delta _{ss^{\prime }}, \\
c_{ns}c_{n^{\prime }s^{\prime }}+c_{n^{\prime }s^{\prime }}c_{ns} &=&\delta
_{n\overline{n}^{\prime }}\delta _{s\overline{s}^{\prime }}\;.
\end{eqnarray}
In terms of the $c_{ns}$ operators the Hamiltonian can be written in the form

\begin{equation}
\hat{H}=\sum_{\{p,n,s\}}c_{ns}^{+}{\cal H}_{(nn^{\prime })(ss^{\prime
})}c_{n^{\prime }s^{\prime }}\;,
\end{equation}
where the summation is performed over all momenta, Nambu and spin indices.
The matrix $\check{{\cal H}}$ is given by

\begin{equation}
\check{{\cal H}}=\frac{1}{2}\left\{ \left[ \left( \xi _{p}\delta
_{pp^{\prime }}+eV\right) +U_{imp}\right] \hat{\tau}_{3}\otimes \hat{\sigma}%
_{0}+\widetilde{\hat{\Delta}}\otimes \hat{\sigma}_{3}-h\left[ \left( \hat{%
\tau}_{0}\otimes \hat{\sigma}_{3}\right) \cos \alpha +\left( \hat{\tau}%
_{3}\otimes \hat{\sigma}_{2}\right) \sin \alpha \right] \right\}
\end{equation}
The matrices $\hat{\tau}_{i}$ and $\hat{\sigma}_{i}$ are the Pauli matrices
in the Nambu and spin space respectively; $i=0,1,2,3$, where $\hat{\tau}_{0}$
and $\sigma _{0}$ are the corresponding unit matrices. We have assumed that
the exchange field ${\bf h}$ has the components ${\bf h}=h(0,\sin \alpha
,\cos \alpha )$; this is the case we consider in the next sections. The
matrix order parameter equals $\widetilde{\hat{\Delta}}=\hat{\tau}%
_{1}Re\Delta -\hat{\tau}_{2}Im\Delta $. Now we can define the matrix Green's
functions (in the Nambu$\otimes $spin space) in the Keldysh representation
in a standard way

\begin{equation}
\check{G}(t_{i},t_{k}^{\prime })=\frac{1}{i}\left\langle T_{C}\left(
c_{ns}(t_{i})c_{n^{\prime }s^{\prime }}^{+}(t_{k}^{\prime })\right)
\right\rangle ,
\end{equation}
where the temporal indices take the values $1$ and $2$, which correspond to
the upper and lower branch of the contour $C$, running from $-\infty $ to $%
+\infty $ and back to $-\infty $. The quasiclassical Green's functions $%
\check{g}(t_{i},t_{k}^{\prime })$ are defined as usual ( Refs.\cite
{Eilen,LOvchQuas})

\begin{equation}
\check{g}({\bf p}_{{\bf F}},{\bf r})=\frac{i}{\pi }\left( \hat{\tau}%
_{3}\otimes \hat{\sigma}_{0}\right) \int d\xi _{p}\check{G}%
(t_{i},t_{k}^{\prime };{\bf p},{\bf r})\;.
\end{equation}
We also introduce, as it was done by Larkin and Ovchinnikov (Ref. \cite
{LOvchBook}) , a hypermatrix $\check{{\bf g}}$ . The matrix elements of $%
\check{{\bf g}}$ are the retarded $\check{g}^{R}$, advanced $\check{g}^{A}$
and the Keldysh $\check{g}^{K}$component. Thus, $\check{{\bf g}}$ has the
form

\begin{equation}
\check{{\bf g}}=\left( 
\begin{array}{cc}
\check{g}^{R} & \check{g}^{K} \\ 
0 & \check{g}^{A}
\end{array}
\right) .  \label{matrix}
\end{equation}
The functions $\check{g}^{R(A)}$ and $\check{g}^{K}$ can be expressed in
terms of the time-ordered Green's functions $\check{g}(t_{i},t_{k})$ as
follows

\begin{eqnarray}
\check{g}^{R(A)} &=&\check{g}(t_{1},t_{1}^{\prime })-\check{g}%
(t_{1(2)},t_{2(1)}^{\prime }) \\
\check{g}^{K} &=&\check{g}(t_{1},t_{2}\check{^{\prime }})+\check{g}%
(t_{2},t_{1}^{\prime }).
\end{eqnarray}
The equation for the hypermatrix $\check{{\bf g}}$ can be easily derived in
the same way as it was done for the case of a superconductor with
spin-independent interactions ( Ref.\cite{LOvchBook}). We are interested in
the diffusive limit. The symmetric component of the matrix $\check{{\bf g}}$
with respect to the momentum direction ${\bf p}_{{\bf F}}$ satisfies the
equation:

\begin{equation}
-iD\nabla \left( \check{{\bf g}}\nabla \check{{\bf g}}\right) +i\left( \hat{
\tau}_{3}\otimes \hat{\sigma}_{0}.\partial _{t}\check{{\bf g}}+\partial
_{t^{\prime }}\check{g}.\hat{\tau}_{3}\otimes \hat{\sigma}_{0}\right) +eV(t) 
\check{{\bf g}}-\check{{\bf g}}eV(t^{\prime })+\left[ \hat{\Delta}\otimes 
\hat{\sigma}_{3},\check{{\bf g}}\right] + \left[\check{M}_{h},\check{{\bf g}}%
\right] =0 \; .  \label{usadel}
\end{equation}
Here $D=vl/3$ is the diffusion coefficient, $\check{M}_{h}=h\left(\hat{\tau}%
_{3}\otimes \hat{\sigma}_{3}\cos \alpha +\hat{\tau}_{0}\otimes \hat{\sigma}%
_{2}\sin \alpha \right)$, and

\[
\hat{\Delta}=-\widetilde{\hat{\Delta}}.\hat{\tau}_{3}=\left( 
\begin{array}{cc}
0 & \Delta  \\ 
-\Delta ^{\ast } & 0
\end{array}
\right) \;.
\]
Eq.The  (\ref{usadel}) is supplemented by the normalization condition 
\begin{equation}
\check{{\bf g}}.\check{{\bf g}}=\check{1}\;,  \label{normalization}
\end{equation}
and by the self-consistency equation 
\begin{equation}
\Delta =\frac{\lambda }{16}{\rm Tr}\left( \hat{\tau}_{1}-i\hat{\tau}%
_{2}\right) \otimes \hat{\sigma}_{3}\int d\epsilon \check{g}^{K}\;.
\label{selfconsist}
\end{equation}
If the magnetization, and hence the exchange field $h$, is constant in the
ferromagnetic layers, the angle $\alpha $ in Eq. (\ref{usadel}) can be
excluded with the help of the following unitary transformation 
\begin{equation}
\widetilde{\check{{\bf g}}}=\check{U}^{+}.\check{{\bf g}}.\check{U}\;,
\label{rotation}
\end{equation}
where $\check{U}=\hat{\tau}_{0}\otimes \hat{\sigma}_{0}\cos (\alpha
/2)+i\sin (\alpha /2)\hat{\tau}_{3}\otimes \hat{\sigma}_{1}$. We consider
F/S structures, therefore we need the boundary conditions for Eq. (\ref
{usadel}) at the interface between the conductors ``$a$'' and ``$b$'' \cite
{Zaitsev} 
\begin{equation}
\check{{\bf g}}_{a}\partial _{x}\check{{\bf g}}_{a}=\frac{\rho _{a}}{2R_{a/b}%
}\left[ \check{{\bf g}}_{a},\check{{\bf g}}_{b}\right] \;,
\label{boundary-condition}
\end{equation}
where $\rho _{a}$ is the specific resistivity of the conductor ``$a$'' and $%
R_{a/b}$ is the interface resistance per unit area. We assumed that there
are no spin-flip processes at the interface. In the presence of
spin-processes at the boundary the condition (\ref{boundary-condition}) can
be generalized (Ref.\cite{Millis}). The current density is determined by the
usual expression 
\begin{equation}
I_{J}=\frac{1}{16\rho }{\rm Tr}\left( \hat{\tau}_{3}\otimes \hat{\sigma}%
_{0}\right) \int {\rm d}\epsilon \left( \check{g}^{R}.\partial _{x}\check{g}%
^{K}+\check{g}^{K}.\partial _{x}\check{g}^{A}\right) \,.  \label{current}
\end{equation}

\clearpage

FIG.1: The S/F/S system.

FIG.2: Dependence of the normalized critical current
  $J_c/\gamma^2\Delta_0$ on the thickness $d$ of the ferromagnet for
  $\Delta_0\tau=0.05$ and $T/\Delta_0=0.01$. The curves for $h\tau=0.1$ and
  $h\tau=0.5$ are multiplied by a factor $10$ for clarity.

FIG.3: Dependence of the normalized critical current
  $J_c/\gamma^2\Delta_0$ on the temperature for $\Delta_0\tau=0.03$, 
$h\tau=0.06$ and  $d/l=\pi$.

FIG.4: Dependence of the normalized critical current
  $J_c/\gamma^2\Delta_0$ on the thickness $d$ of the ferromagnet for
  $\Delta_0\tau=0.05$ and $T/\Delta_0=0.1$.

FIG.5: Dependence of the normalized critical current
  $J_c/\gamma^2\Delta_0$ on the temperature for $d/l=1$. The two upper curves
  correspond to the case $h\tau=5$, and the the two lower curves to the case
  $h\tau=10$. For clarity, in the three lower curves, the values of the
  critical current have been  multiplied by a factor $5$.

FIG. 6: The dependence of the critical current on $T$ for $h\tau=0.06$,
  $\Delta_0\tau=0.03$, $d/l=\pi$ and different values of $Ql$.

FIG.7: SF/I/SF system.

FIG.8: S/F bilayer.

FIG.9: Spatial dependence of the current density. Here
  $\tilde{J}=\frac{j}{\gamma^2\Delta_0\sigma}\frac{\phi_0}{Bl}$, $d/l=10$,
  $T/\Delta_0=0.1$ and $\Delta_0\tau=0.05$. For clarity, the values of the
  current density for $h\tau=0.1$ have been multiplied by a factor 5.

FIG.10: Dependence of the  total current
  on the thickness $d$ of the ferromagnet film for
  $\Delta_0\tau=0.05$ and $T/\Delta_0=0.1$. Here $\tilde{I}=\frac{I}{\gamma^2\Delta_0\sigma}\frac{\phi_0}{Bl^2}$

FIG.11: Dependence of the total current
   on the parameter $h\tau$ for  $\Delta_0\tau=0.05$ and
   $T/\Delta_0=0.1$. Here
   $\tilde{I}=\frac{I}{\gamma^2\Delta_0\sigma}\frac{\phi_0}{Bl^2}$. At
   $h\tau=0$: $\tilde{I}(0)=-34,-25,-18$ for $d/l=1,5,10$ respectively.

\end{document}